\documentclass[11pt]{article}

\usepackage{deauthor}
\usepackage{times}
\usepackage{paralist} 
\usepackage{verbatim} 
\usepackage[pdftex]{graphicx}
\usepackage{subfigure}
\usepackage{breakurl}
\usepackage{color}
\usepackage{multirow}
\usepackage{listings}
\usepackage{array}
\usepackage{subfigure}
\usepackage{graphicx}
\usepackage{tabularx}
\usepackage{lscape}
\usepackage{rotating}
\usepackage{algorithm}
\usepackage{algorithmic}
\usepackage{multicol}
\usepackage{xspace}
\usepackage{amsfonts}
\usepackage{alltt,cite}
\usepackage[normalem]{ulem}

\usepackage{fancyheadings}
\usepackage[pdftex]{hyperref}

\newcommand{\dd}{DeepDive\xspace}
\newcommand{\pdd}{PaleoDeepDive\xspace}

\def\compactify{\itemsep=0pt \topsep=0pt \partopsep=0pt \parsep=0pt}

\begin{document}

%\title{DeepDive: Building a Trained System for \\Knowledge Base Construction}
\title{Feature Engineering for Knowledge Base Construction}
\author{Christopher R\'{e}$^\dagger$~~~~Amir Abbas Sadeghian$^\dagger$~~~~Zifei Shan$^\dagger$\\
Jaeho Shin$^\dagger$~~~~Feiran Wang$^\dagger$~~~~Sen Wu$^\dagger$~~~~Ce Zhang$^{\dagger\ddagger}$\\
$^\dagger$Stanford University\\
$^\ddagger$University of Wisconsin-Madison\\
\{chrismre, amirabs, zifei, jaeho.shin, feiran, senwu, czhang\}@cs.stanford.edu
}

\maketitle

\begin{abstract}
Knowledge base construction (KBC) is the process of populating a
knowledge base, i.e., a relational database together with inference
rules, with information extracted from documents and structured
sources. KBC blurs the distinction between two traditional database
problems, information extraction and information integration. For the
last several years, our group has been building knowledge bases with
scientific collaborators. Using our approach, we have built knowledge
bases that have comparable and sometimes better quality than those
constructed by human volunteers. In contrast to these knowledge bases,
which took experts a decade or more human years to construct, many of
our projects are constructed by a single graduate student.

Our approach to KBC is based on joint probabilistic inference and
learning, but we do not see inference as either a panacea or a magic
bullet: inference is a tool that allows us to be systematic in how we
construct, debug, and improve the quality of such systems. In
addition, inference allows us to construct these systems in a more
loosely coupled way than traditional approaches. To support this idea,
we have built the DeepDive system, which has the design goal of
letting the user ``think about features---not algorithms.'' We think
of DeepDive as declarative in that one specifies what they want but
not how to get it. We describe our approach with a focus on feature
engineering, which we argue is an understudied problem relative to its
importance to end-to-end quality.
\end{abstract}

\section{Introduction}

This document highlights what we believe is a critical and
underexplored aspect of building high-quality knowledgebase
construction (KBC) systems: ease of feature engineering. The
hypothesis of our work is that the easier the feature engineering
process is to debug and diagonse, the easier it is to improve the
quality of the system. The single most important design goal of \dd is
to make KBC systems easier to debug and improve. Although such
techniques are important for end-to-end quality, techniques to debug
KBC systems are understudied.

To describe our thoughts more precisely, we describe our motivation
for building KBC systems, our choice of language that defines the
syntax of our feature engineering problem, and a set of debugging
techniques that we have found useful:\footnote{This document is
  superficial, and we refer the interested reader to more detailed
  material including example code and data that are available
  online. \url{http://deepdive.stanford.edu} contains the latest
  material and is under active construction.}

\begin{description}
\item \textbf{Motivation.} In Section~\ref{sec:kbc}, we discuss a key
  motivating application for \dd, KBC. We take a holisitic approach
  that performs integration and extraction as a single task (rather
  than as two separate tasks).  This holistic approach allows is to
  acquire data at all levels, including from larger and more diverse
  corpora, more human annotations, and/or larger (but still incomplete)
  knowledge bases. In turn, we can choose the source of data that has
  the best available signal for the least cost, a concept that we
  call {\em opportunistic acquisition}.

\item \textbf{Language.} In Section~\ref{sec:overview}, we describe
  our choice of language model with the goal of abstracting the
  details of statistical learning and inference from the user to
  the extent possible. To integrate domain knowledge and support
  standard tools on top of \dd, we built our system on the relational
  model. Originally, we had our own custom-built language for
  inference based on Markov
  Logic~\cite{DBLP:journals/ml/RichardsonD06}, but we found that this
  language was unfamiliar to scientists. Instead, we decided to be as
  language agnostic as possible. A scientist can now write in almost
  any language to engineer features in \dd. However, for scalability
  reasons, bulk operations are written in SQL.

\item \textbf{Debugging. } In Section~\ref{sec:error_analysis}, we
  describe our approach to debugging. A key challenge is how to
  identify the relevant domain knowledge to integrate and what types
  of input sources have valuable signal.  In our experience, without a
  systematic error analysis, it is not uncommon for developers to add
  rules that do not significantly improve the quality, which we call
  {\em prematurely optimizing} the KBC system. This is analogous to
  the popular engineering mistake of optimizing code without first
  profiling its end-to-end performance, akin to a failure to
  appreciate Amdahl's law. We view the key contribution of this work
  as highlighting the importance of error analysis in KBC
  applications.
\end{description}

\noindent
Performance is also a major challenge. In our KBC systems using \dd,
we may need to perform inference and learning on billions of highly
correlated random variables. Therefore, one of our technical focus
areas has been to speed up probabilistic
inference~\cite{Zhang:2014:VLDB,
  Zhang:2013:SIGMOD,Niu:2011:NIPS,Niu:2011:VLDB,Niu:2012:ICDM}. Although
this challenge is important, we do not focus on it in this paper, but
there is much interesting work in this
direction~\cite{Canny:2013:KDD,Jampani:2008:SIGMOD,Cai:2013:SIGMOD,Wang:2011:SIGMOD,Sen:2009:VLDBJ,DBLP:conf/sigmod/ChenW14}
including work on how to automatically select algorithms from a
declarative specification~\cite{DBLP:conf/cidr/KraskaTDGFJ13,DBLP:journals/corr/abs-1108-0294,DBLP:journals/corr/abs-1108-0294,Niu:2012:ICDM}

\begin{figure*}[t]
\centerline{\includegraphics[width=0.99\textwidth]{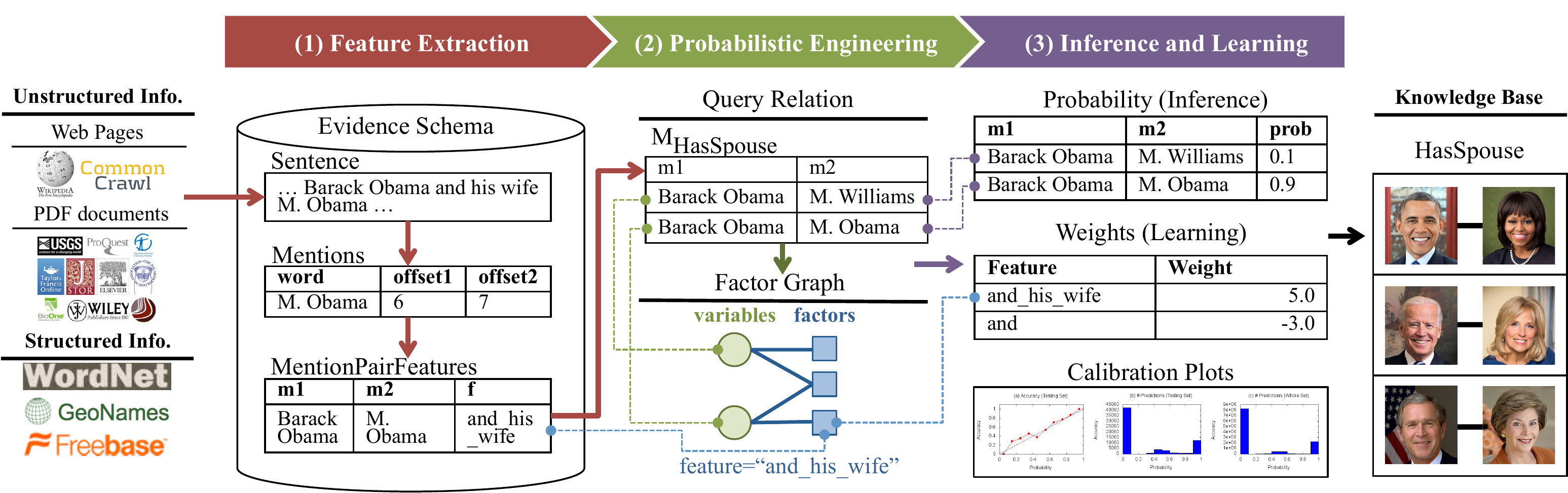}}
\caption{An overview of a KBC system built with \dd
that takes as input both structured and unstructured
information and creates a knowledge base as the output.
There are three phases of \dd's execution model:
(1) Feature Extraction; (2) Probabilistic Engineering; and (3)
Inference and Learning. Section~\ref{sec:overview}
contains a more detailed walkthrough of these phases, and
Figure~\ref{fig:overview} shows
more details of how to conduct tasks in these three phases
using SQL and script languages, e.g., Python.}
\label{fig:ddoverview}
\end{figure*}

\begin{figure*}
\centering
\centerline{\includegraphics[width=0.99\textwidth]{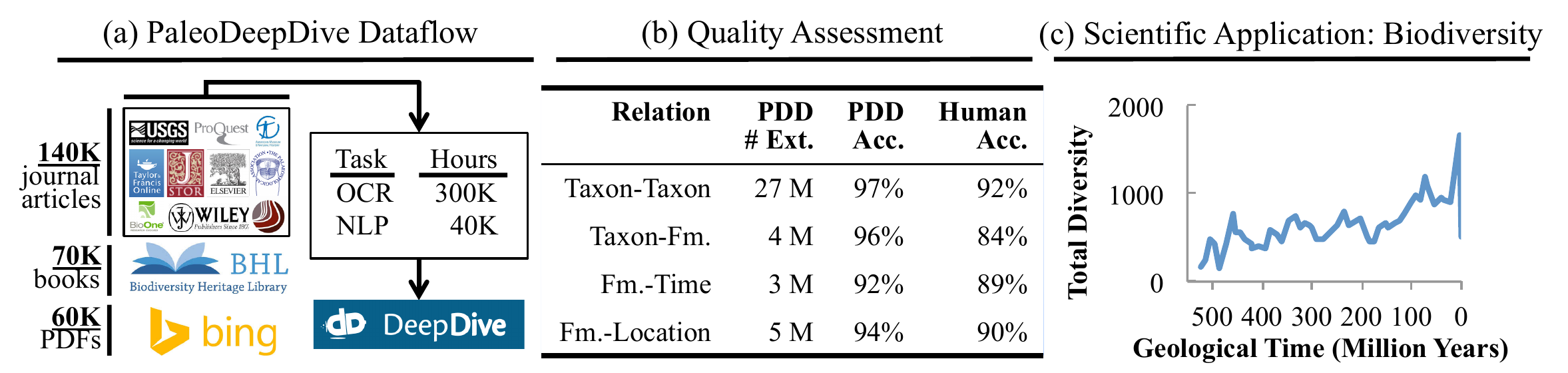}}
\caption{\pdd, a KBC system for paleontology built with \dd.
(a) The dataflow of \pdd; (b) The assessment of quality
for extractions in \pdd (PDD); (c) One scientific application
of \pdd---biodiversity during the Phanerozoic period.
More detailed results can be found in Peters et al.~\cite{Peters:2014:ArXiv}.}
\label{fig:quality}
\end{figure*}

\section{Knowledge Base Construction: The Case for Opportunistic Acquisition}\label{sec:kbc}

Knowledge base construction is the process of populating a knowledge
base with facts extracted from text, tabular data expressed in text
and in structured forms, and even maps and figures. In {\it
  sample-based science}~\cite{Peters:2014:ArXiv}, one typically
assembles a large number of facts (typically from the literature) to
understand macroscopic questions, e.g., about the amount of carbon in
the Earth's atmosphere throughout time, the rate of extinction of
species, or all the drugs that interact with a particular gene. To
answer such questions, a key step is to construct a high-quality
knowledge base, and some forward-looking sciences have undertaken
decade-long sample collection efforts, e.g., PaleoDB.org and
PharmaGKB.org.

In parallel, KBC has attracted interest from
industry~\cite{Ferrucci:2010:AI,Zhu:2009:WWW} and
academia~\cite{Krishnamurthy:2009:SIGMODRec,
  Shen:2007:VLDB,Suchanek:2009:WWW,Nakashole:2011:WSDM,Etzioni:2004:WWW,
  Yates:2007:NAACL,Banko:2007:IJCAI,Kasneci:2009:SIGMODR,Betteridge:2009:AAAI,
  Carlson:2010:AAAI,Poon:2007:AAAI,Niu:2012:IJSWIS}. To understand the
common patters in KBC systems, we are actively collaborating with
scientists from a diverse set of domains, including
geology~\cite{Zhang:2013:SIGMODDEMO},
paleontology~\cite{Peters:2014:ArXiv}, pharmacology for drug
repurposing, and others. We briefly describe an application that we
have constructed.

\begin{example}[Paleontology and KBC~\cite{Peters:2014:ArXiv}]
Paleontology is based on the description and biological classification
of fossils, an enterprise that has played out in countless collecting
expeditions, museum visits, and an untold number of scientific
publications over the past four centuries. One central task for
paleontology is to construct a knowledge base about fossils from
scientific publications, and an existing knowledge base compiled by
human volunteers has greatly expanded the intellectual reach of
paleontology and led to many fundamental new insights into
macroevolutionary processes and the nature of biotic responses to
global environmental change.  However, the current process of using
human volunteers is usually expensive and time-consuming. For example,
PaleoDB, one of the largest such knowledge bases, took more than 300
professional paleontologists and 11 human years to build over the last
two decades, resulting in \url{PaleoDB.org}. To get a sense of the
impact of this database on this field, at the time of writing, this
dataset has contributed to 205 publications, of which 17
have appeared in {\em Nature} or {\em Science}.

This provided an ideal test bed for our KBC research. In particular,
we constructed a prototype called
PaleoDeepDive~\cite{Peters:2014:ArXiv} that takes in PDF documents. As
a result, this prototype attacks challenges in optical character
recognition, natural language processing, and information extraction
and integration. Some statistics about the process are shown in
Figure~\ref{fig:quality}(a). As part of the validation of this system,
we performed a double-blind experiment to assess the quality of the
system versus the PaleoDB. We found that the KBC system built on \dd
has achieved comparable---and sometimes better---quality than a
knowledge base built by human volunteers over the last
decade~\cite{Peters:2014:ArXiv}. Figure~\ref{fig:quality}(b)
illustrates the accuracy of the results in PaleoDeepDive.
\end{example}

\begin{figure*}
\centering
\centerline{\includegraphics[width=0.99\textwidth]{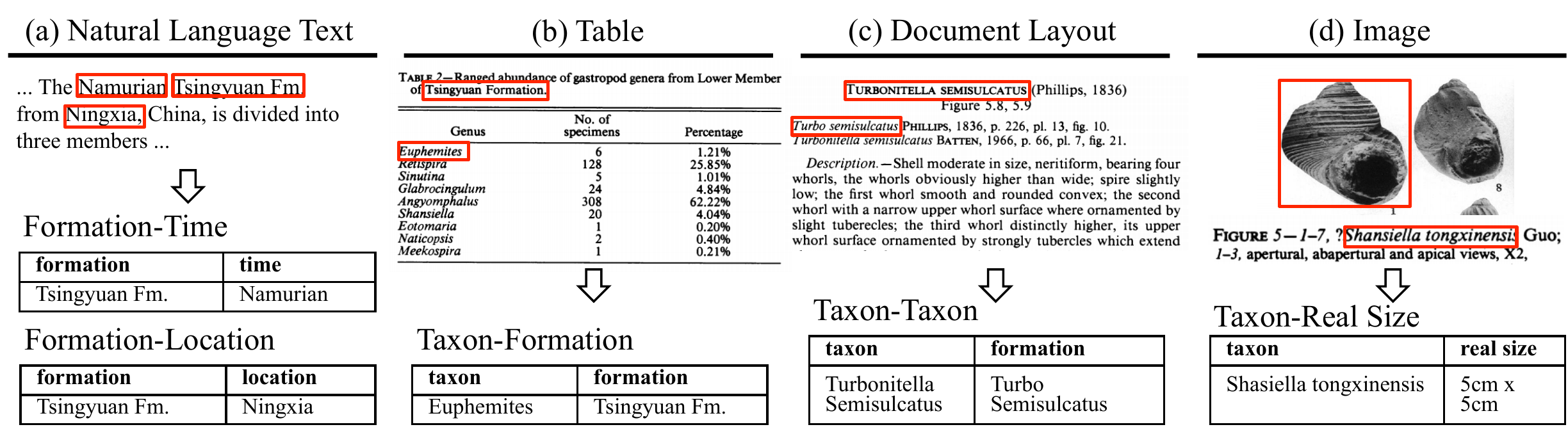}}
\caption{An illustration of data sources that \pdd
supports for KBC.}
\label{fig:sources}
\end{figure*}

We have found that text is often not enough: often, the data that are
interesting to scientists are located in the tables and figures of
articles. For example, in geology, more than 50\% of the facts that we
are interested in are buried in
tables~\cite{Govindaraju:ACL:2012}. For paleontology, the relationship
between taxa, as known as taxonomy, is almost exclusively expressed in
section headers~\cite{Peters:2014:ArXiv}.  For pharmacology, it is not
uncommon for a simple diagram to contain a large number of metabolic
pathways. To build a KBC system with the quality that scientists will
be satisfied with, we need to deal with these diverse sources of
input.  Additionally, external sources of information (other knowledge
bases) typically contain high-quality signals (e.g.,
Freebase\footnote{\url{http://www.freebase.com/}} and
Macrostrat\footnote{\url{http://macrostrat.org/}}).  Leveraging these
sources in information extraction is typically not studied in the
classical information extraction context. To perform high-quality and
high-coverage knowledge extraction, one needs a model that is able to
ingest whatever presents itself {\em opportunistically}---that is, it
is not tied solely to text but can handle more general
extraction and integration.

\paragraph*{Opportunistic Acquisition.} 
We outline why we believe that the system must be able to {\it
  opportunistically acquire} data, by which we mean {\it the ability
  to acquire data from a wide variety of sources only to the degree to
  which they improve the end-to-end result quality}.

We describe the mechanisms by which we have improved the quality in
KBC applications.\footnote{A complete list of the features we used
  for PaleoDeepDive can be found in our
  paper~\cite{Peters:2014:ArXiv}.} Consider one pattern that we
observed in each of these KBC systems: we start from a
state-of-the-art relation extraction model, e.g., a logistic
regression model with rich linguistic
features~\cite{Mintz:2009:ACL,Hoffmann:2010:ACL}. Out of the box, such
models do not have the quality that a domain expert requires. As a
result, the expert must improve the model. This process of improving
the model accounts for the dominant amount of time in which they
interact with the system. To improve the extraction quality in terms
of both precision and recall, the system needs some additional domain
knowledge. For example, paleontologists might want to add a rule to
improve the recall of a KBC system by informing the system about
geological time; e.g., if a fossil appears in Late Cretaceous, then it
also appears in the Cretaceous period.  Biologists might want to add a
rule to improve precision stating that one mention should only be
recognized as a certain class of drug if there is clear linguistic
clue that the drug is actually administered to a patient. These rules,
however, must be expressed by domain scientists explicitly for the KBC
system. Allowing users to express their knowledge is a challenge that
has been studied in the extraction literature for some time, notably
by SystemT~\cite{Krishnamurthy:2009:SIGMODRec,Li:2011:ACL}, which
allows users to express their knowledge using declarative queries. In
the next section, we describe our joint or collective approach, which
is combined with a powerful technique called {\it distant supervision}
that has allowed us to build knowledge bases with low cost. This
technique allows \dd to take in this knowledge in many ways: rules,
features, example datasets, and labeled data.

Choosing how to improve the system in the most effective way is
arguably the key pain point in KBC. Nevertheless, we have noticed that
there is a tendency to {\it prematurely optimize the quality} of
intermediate results if the system is constructed in the absence of
end-to-end goals. In our experience, we found that many of these
premature optimizations seem reasonable in isolation and do locally
improve the quality of the system. However, such optimizations have
only marginal improvement on the end-to-end quality. As there is a
never ending see of intermediate fixes prioritizing these fixes seems
to be the critical issue. As a result, we advocate that one should
prioritze how to improve the quality of the system based on how it
affects the end-to-end quality of the system. We describe our first
cut of how to make these choices in Section~\ref{sec:error_analysis},
with an emphasis on avoiding premature optimization.

% In this paper, we describe the \dd approach
% with respect to the above three challenges.
% In Section~\ref{sec:overview}, we describe
% the data model and language of \dd, and
% given an example using a system we built called
% \pdd. In Section~\ref{sec:error_analysis},
% we describe our workflow of error analysis
% and how to improve a KBC program. In Section~\ref{sec:performance},
% we briefly discuss techniques that we deployed
% to improvement the performance of \dd.
% We survey related work in Section~\ref{sec:related_work}
% and discuss future directions in Section~\ref{sec:future}.

% \begin{figure*}[t]
% \centerline{\includegraphics[scale=0.6]{figures/overview.pdf}}
% \caption{The Overview of a KBC System Built with \dd.
% The input for the KBC system is the corpora and resources
% stored as relations in RDBMS. A \dd program contains two
% components: (1) a set of feature extractors that populate
% new relations; and (2) a set of inference rules that specify
% a factor graph. Given a factor graph grounded using a \dd
% program, \dd conducts (3) statistical inference and learning,
% and (4) reports statistics in the form of calibration plots.}
% \label{fig:overview}
% \end{figure*}

\begin{figure*}[t]
\centerline{\includegraphics[width=0.99\textwidth]{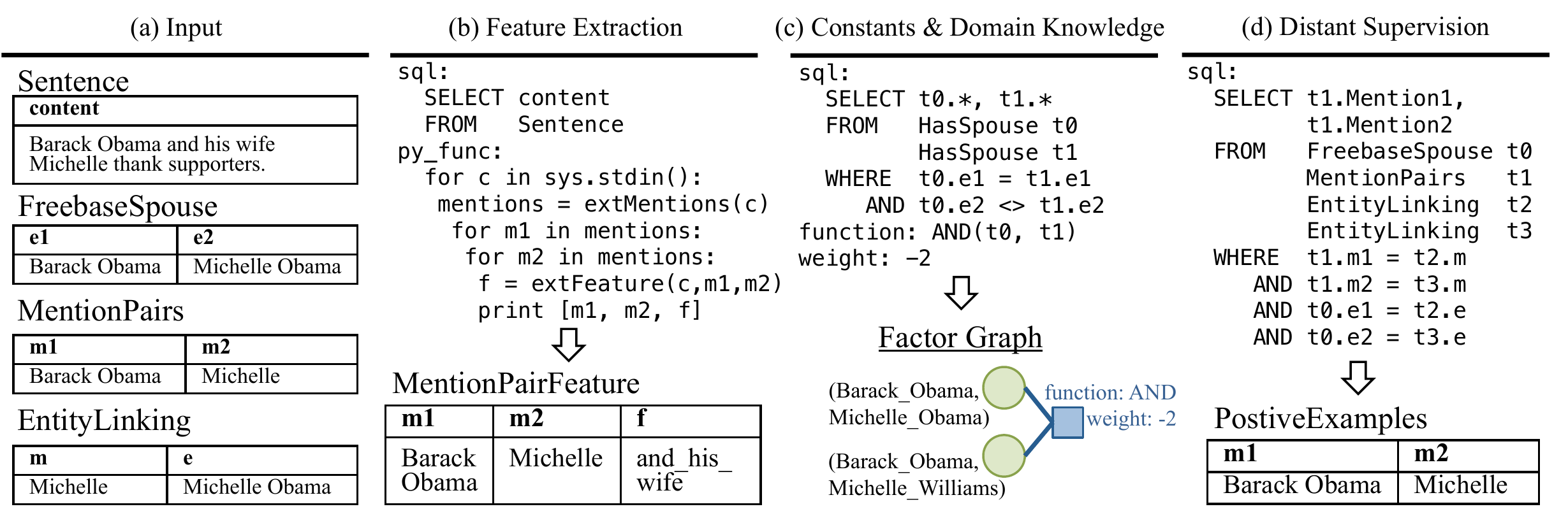}}
\caption{Illustration of popular operations in \dd. (a) Prepare
data sets in relational form that can be used by \dd. (b) Generate
labels using distant supervision with SQL; (c) Integrate 
constraints with SQL and logic functions; (d) Extract
features with SQL and script languages (e.g., Python).}
\label{fig:overview}
\end{figure*}

\section{A Brief Overview of \dd} \label{sec:overview}

\begin{figure*}[t]
\centerline{\includegraphics[width=1.05\textwidth]{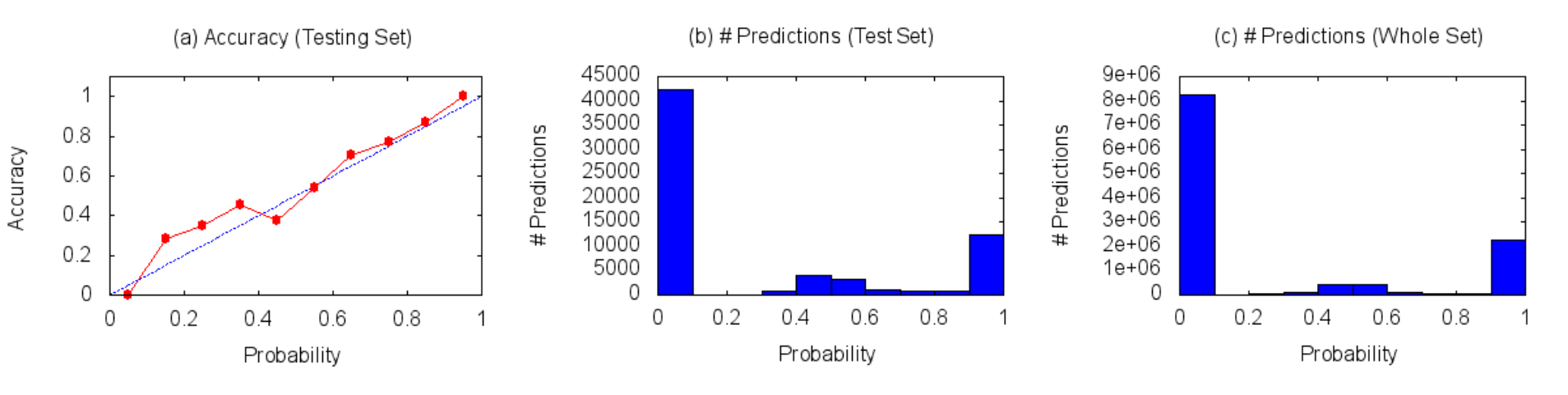}}
\caption{Illustration of calibration plots in \dd.
%\yell{Testing set to Test Set}
}
\label{fig:cali}
\end{figure*}

We briefly introduce the programming and execution model of
\dd. Figure~\ref{fig:overview} shows the code that a user writes to
interact with \dd.  The reader can find a more detailed description of
the language model of \dd in our previous
work~\cite{Niu:2012:IJSWIS,Zhang:2013:SIGMOD} and on \dd's Web site.
There are three phases in \dd's execution model, as shown in
Figure~\ref{fig:ddoverview}:

\begin{description}
\item \textbf{(1) Feature Extraction.} The input to this stage often
  contains both structured and unstructured information, and the goal
  is to produce a relational database that describes the features or
  signals of the data, which we call the {\em evidence schema}. We use
  the phrase evidence to emphasize that, in this stage,
  data is not required to be precisely correct as in traditional ETL;
  as a result, this is a much lighter ETL process.

\begin{itemize}
\item One responsibility of this phase is to run various OCR and NLP
  tools to acquire information from text, HTML, or images. The output
  database is often unnormalized: it may contain JSON objects to
  represent parse trees or DOM structures. This phase is essentially a
  high-throughput workflow system and may involve MapReduce jobs,
  Condor jobs, and SQL queries.

\item The user also performs feature extraction in that they write
  user-defined functions (UDF) over existing query and evidence
  relations. \dd supports both ways, and the user can use SQL queries,
  and script languages, e.g., Python or Perl, to specify
  UDFs. Figure~\ref{fig:overview}(b) and (d) show two examples.
\end{itemize}

\item \textbf{(2) Probabilistic Engineering.} The goal of this phase is
  to transform the evidence schema into a probabilistic model,
  specifically a factor graph that specifies:

\begin{itemize}
\item The set of random variables that the user wants to model. To
  define the set of random variables in \dd is to create new relations
  called query relations, in which each tuple corresponds to one random
  variable. This operation can be done by using SQL queries on
  existing relations.

\item How those random variables are correlated; e.g., ``The mention
  `Obama' refers to the president is correlated with the random
  variable that indicates whether `Obama' is a person.'' To specify
  {\em how}, the user specifies a factor function.
  Figure~\ref{fig:overview}(c) shows a rough example that describes
  the intuition that {\it ``people tend to be married to only a single
    person.''} One (of many ways) to say this is to say that there is
  some correlation between pairs of married tuples; i.e., using the
  logical function \textsf{AND(t0, t1)} returns 1 in possible worlds
  in which both married tuples are true and 0 in others.  \dd then
  learns the ``strength'' of this correlation from the data, which is encoded
  as weight.\footnote{A weight is roughly the log odds, i.e., the
    $\log \frac{p}{1-p}$ where $p$ is the marginal probability of this
    random variable. This is standard in Markov Logic
    Networks~\cite{DBLP:journals/ml/RichardsonD06}, on which much of
    \dd's semantics are based.} Here, $-2$ indicates that it is
  less likely that both married tuples are correct. This phase is also
  where the user can write logical constraints, e.g., hard functional
  constraints.
\end{itemize}

Our previous work has shown that this grounding phase~\cite{Niu:2011:VLDB} can
be a serious bottleneck if one does not use scalable relational
technology. We have learned this lesson several times.

\item \textbf{(3) Inference and Learning.} This phase is largely opaque
  to the user: it takes the factor graph as input, estimates the
  weights, performs inference, and produces the output database along
  with a host of diagnostic information, notably calibration plots
  (see Fig.~\ref{fig:cali}). More precisely, the output of \dd is a
  database that contains each random variable declared by the user
  with its marginal probability.  For example, one tuple in the
  relation \textsf{HasSpouse} might be (Barack Obama, Michelle Obama),
  and ideally, we hope that \dd outputs a larger probability for this tuple
  as output.
\end{description}

\subsection{Operations to Improve Quality in DeepDive}

\begin{figure*}[t]
\centerline{\includegraphics[width=0.99\textwidth]{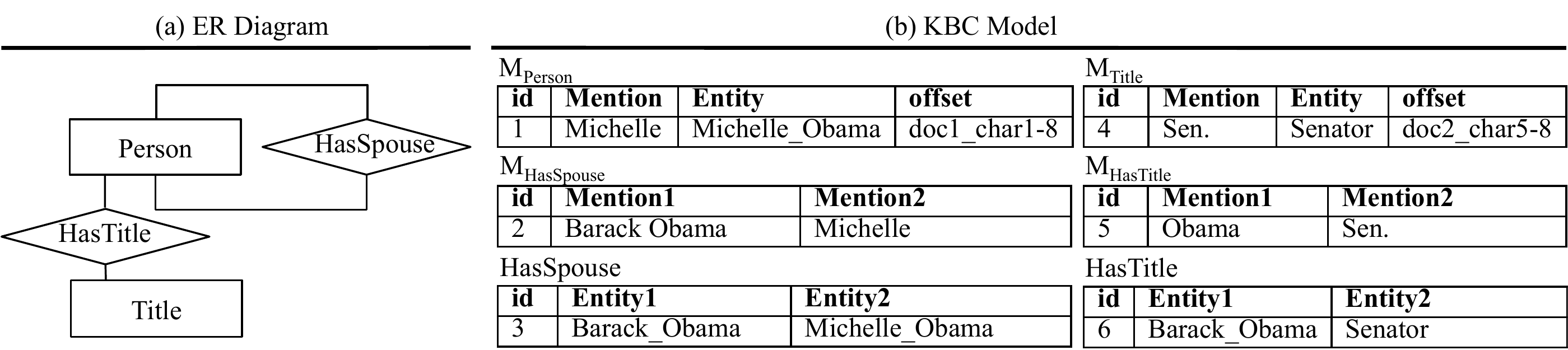}}
\caption{Illustration of the KBC model in \dd.  (a) An example ER
  diagram for the TAC-KBP application. (b) The corresponding KBC
  model, where each tuple in each relation corresponds to one random
  variable that \dd will predict. The column ``offset'' in mention
  tables $M_{Person}$ and $M_{Title}$ is the pointer to the position
  in the text, e.g., character offset in a document, and might be
  more complicated in real applications, e.g., a bounding box in
  a PDF.}
\label{fig:kbc}
\end{figure*}
We illustrate the KBC model using one application called
TAC-KBP,\footnote{http://www.nist.gov/tac/2013/KBP/} in which the
target knowledge base contains relations between persons, locations,
and organizations. As is standard in the information extraction
literature, a mention is a sequence of tokens in text, while an entity
refers to the real-world object in the
database. Figure~\ref{fig:kbc}(a) shows an excerpt from the ER
diagram of the database. In this diagram, we have two types of
entities, \textsf{Person} and \textsf{Title}, and two relations,
\textsf{HasSpouse} and \textsf{HasTitle}.  Figure~\ref{fig:kbc}(b)
shows the KBC model induced from this ER diagram. For example,
consider the \textsf{Person} entity and the \textsf{HasSpouse}
relation.  For \textsf{Person}, the KBC model contains a relation
$M_{Person}$, which contains the candidate linking between a person
mention (e.g., Michelle) to the person entity (e.g., Michelle\_Obama
or Michelle\_Williams).  The relation $M_{HasSpouse}$ contains mention
pairs, which are candidates that participate in the
\textsf{HasSpouse} relation, and the relation $HasSpouse$ contains the
entity pairs. Each tuple of these relations corresponds to a Boolean
random variable.

\paragraph*{Routine Operations in \dd.} 
We describe three routine tasks that a user performs to improve a
KBC system.

\subparagraph*{An Example of Feature Extraction.} 
The concept of a feature is one of the most important concepts for
machine learning systems, and \dd's data model allows the user to use
any scripting language for feature extraction.
Figure~\ref{fig:overview}(b) shows one such example using Python. One
baseline feature that is often used in relation extraction systems is
the word sequence between mention pairs in a
sentence~\cite{Mintz:2009:ACL,Hoffmann:2010:ACL}, and
Figure~\ref{fig:overview}(b) shows an example of extracting this
feature. The user first defines the input to the feature extractor
using an SQL query, which selects all available sentences. Then the
user defines a UDF that will be executed for each tuple returned by
the SQL query.  In this example, the UDF is a Python function that
first reads a sentence from STDIN, extracts mentions from the
sentence, extracts features for each mention pair, and outputs the
result to STDOUT.  \dd will then load the output of this UDF to the
\textsf{MentionPairFeature} relation.

% To tell \dd how to use this feature in probabilistic inference, we can
% write a rule similar to Figure~\ref{fig:overview}(b): \yell{SEE THE
%   ORDER IN THE FIGURE IS WRONG HERE TOO!!}
% \begin{alltt}\small
%   sql:
%     SELECT t0.*, t1.f as f
%     FROM   MentionPairs t0, MentionPairFeatures t1
%     WHERE  t0.m1=t1.m1 AND t0.m2=t1.m2
%   function: Equal(t0, True)
%   weight: ?(f)
% \end{alltt}
% \noindent
% The ``?(feat)'' syntax means that the weight is a function of the
% column ``f'' in the view defined by the SQL query, and will be learned
% automatically by \dd given training data.  \yell{This makes no
%   sense... what does t0 true mean from an application perepsective the
%   types don't match.}

\subparagraph*{Constraints and Domain Knowledge.} One
way to improve a KBC system is to integrate domain knowledge, as we
mentioned in Section~\ref{sec:kbc}. \dd supports this operation by
allowing the user to integrate constraints and domain knowledge as
correlations among random variables, as shown in
Figure~\ref{fig:overview}(c).

Imagine that the user wants to integrate a simple rule that says ``one
person is likely to be the spouse of only one person.''  For example,
given a single entity ``Barack\_Obama,'' this rule gives positive
preference to the case where only one of (Barack\_Obama,
Michelle\_Obama) and (Barack\_Obama, Michelle\_Williams) is true.
Figure~\ref{fig:overview}(c) shows one example of implementing this
rule. The SQL query in Figure~\ref{fig:overview}(c) defines a view in
which each tuple corresponds to two relation candidates with the same
first entity but different second entities.  The function
\textsf{AND(t0, t1)} defines the ``type of correlation'' among
variables, and the weight ``-2'' defines the strength of the
correlation.  This rule indicates that the it is less likely that both
(Barack\_Obama, Michelle\_Obama) and (Barack\_Obama,
Michelle\_Williams) are true (i.e., when \textsf{AND(t0, t1)} returns
1). Typically, \dd is used to learn the weights from
data.

\subparagraph*{Distant Supervision.} One challenge
with building a machine learning system for KBC is generating training
examples. As the number of predicates in the system grows, specifying
training examples for each relation is tedious and expensive. One
common technique to cope with this is distant supervision. Distant
supervision starts with an (incomplete) entity-level knowledge base
and a corpus of text. The user then defines a (heuristic) mapping
between entities in the database and text. This map is used to
generate (noisy) training data for mention-level
relations~\cite{Mintz:2009:ACL,Hoffmann:2010:ACL}. We illustrate this
procedure by example.

\begin{example}[Distant Supervision]
Consider the mention-level relation $M_{HasSpouse}$. To find training
data, we find sentences that contain mentions of pairs of entities
that are married, and we consider the resulting sentences positive
examples. Negative examples could be generated from pairs of persons
who are, say, parent-child pairs. Ideally, the patterns that occur
between pairs of mentions corresponding to mentions will contain
indicators of marriage more often than those that are parent-child
pairs (or other pairs). Selecting those indicative phrases or
features allows us to find features for these relations and generate
training data. Of course, engineering this mapping is a difficult task
in practice and requires many iterations.

Concretely, Figure~\ref{fig:overview}(d) shows an example of distant
supervision in \dd. The \textsf{FreebaseSpouse} relation is an
entity-level relation in Freebase (it contains pairs of married
people). The \textsf{EntityLinking} relation specifies a mapping from
entities to mentions in the text. The user provides an SQL query like
the one shown in Figure~\ref{fig:overview}(d) to produce another
relation, \textsf{PositiveExamples}, that contains mention-level
positive training examples. In our applications, users may spend time
improving this mapping, which can lead to higher quality (but
imperfect) training data much more cheaply than having a human label
this data.
\end{example}

As we have described, the user has at least the above three ways to
improve the system and is free to use one or a combination of them to
improve the system's quality.  The question we address next is, {\it
  ``What should the user do next to get the largest quality
  improvement in the KBC system?''}

\section{Debugging and Improving a KBC System} \label{sec:error_analysis}

A \dd system is only as good as its features and rules. In the last
two years, we have found that understanding which features to add is
the most critical---but often the most overlooked---step in the
process. Without a systematic analysis of the errors of the system,
developers often add rules that do not significantly improve their KBC
system, and they settle for suboptimal quality.  In this section, we
describe our process of error analysis, which we decompose into two
stages: a macro-error analysis that is used to guard against
statistical errors and gives an at-a-glance description of the system
and a fine-grained error analysis that actually results in new
features and code being added to the system.

\subsection{Macro Error Analysis: Calibration Plots}

In \dd, {\em calibration plots} are used to summarize the overall
quality of the KBC results.  Because \dd uses a joint probability
model, each random variable is assigned a marginal probability.
Ideally, if one takes all the facts to which \dd assigns a probability
score of $0.95$, then 95\% of these facts are correct. We believe that
probabilities remove a key element: the developer reasons about
features, not the algorithms underneath. This is a type of {\em
  algorithm independence} that we believe is critical.

DeepDive programs define one or more test sets for each relation,
which are essentially a set of labeled data for that particular
relation. This set is used to produce a calibration
plot. Figure~\ref{fig:cali} shows an example calibration plot for the
\textsf{Formation-Time} relation in PaleoDeepDive, which
provides an aggregated view of how the KBC system behaves. By reading
each of the subplots, we can get a rough assessment of 
the next step to improve our KBC system. We explain each
component below.

As shown in Figure~\ref{fig:cali}, a calibration plot contains three
components: (a) accuracy, (b) \# predictions (test set), which
measures the number of extractions in the test set with a certain
probability; and (c) \# predictions (whole set), which measures the
number of extractions in the whole set with a certain probability. The
test set is assumed to have labels so that we can measure accuracy,
while the whole set does not.

\subparagraph*{(a) Accuracy.} 
To create the accuracy histogram, we bin each fact extracted by
DeepDive on the test set by the probability score assigned to each
fact; e.g., we round to the nearest value in the set $k/10$ for
$k=1,\dots,10$. For each bin, we compute the fraction of those
predictions that is correct. Ideally, this line would be on the
(0,0)-(1,1) line, which means that the \dd-produced probability value
is calibrated, i.e., it matches the {\em test-set
  accuracy}. For example, Figure~\ref{fig:cali}(a) shows a curve for
calibration.  Differences in these two lines can be caused by noise in
the training data, quantization error due to binning, or sparsity in
the training data.

\subparagraph*{(b) \# Predictions (Testing Set).}
We also create a histogram of the number of predictions in each bin.
In a well-tuned system, the \# Predictions histogram should have a
``U'' shape. That is, most of the extractions are concentrated at high
probability and low probability. We do want a number of
low-probability events, as this indicates \dd is considering plausible
but ultimately incorrect alternatives. Figure~\ref{fig:cali}(b) shows
a U-shaped curve with some masses around 0.5-0.6. Intuitively, this
suggests that there is some hidden type of example for which the
system has insufficient features. More generally, facts that fall into
bins that are not in (0,0.1) or (0.9,1.0) are candidates for
improvements, and one goal of improving a KBC system is to ``push''
these probabilities into either (0,0.1) or (0.9,1.0).  To do this, we
may want to sample from these examples and add more features to
resolve this uncertainty.

\subparagraph*{(c) \# Predictions (Whole Set).}
The final histogram is similar to Figure~\ref{fig:cali}(b), but
illustrates the behavior of the system, for which we do not have any
training examples. We can visually inspect that
Figure~\ref{fig:cali}(c) has a similar shape to (b); If not, this
would suggest possible overfitting or some bias in the selection of the
hold-out set.

\begin{figure*}[t]
\centerline{\includegraphics[width=1.05\textwidth]{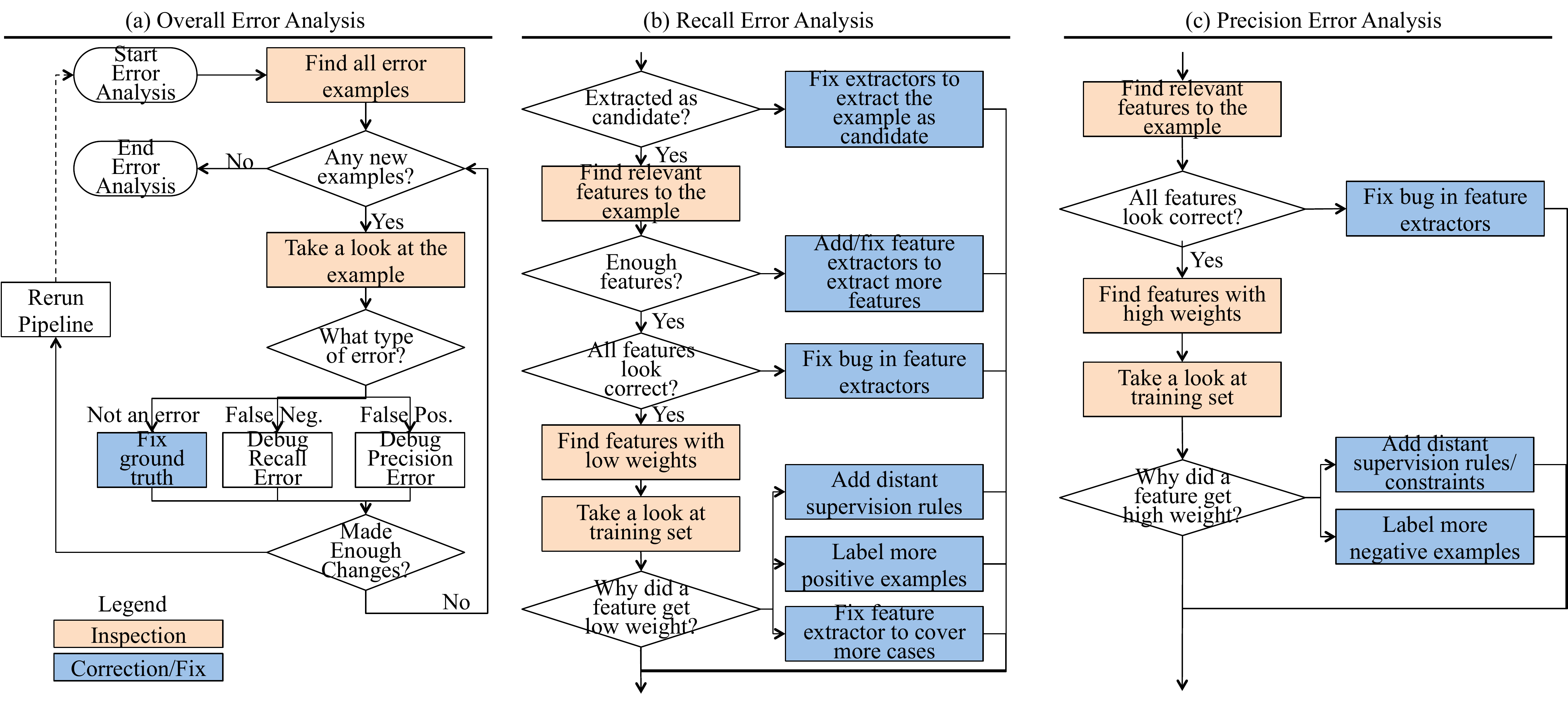}}
\caption{Error Analysis Workflow of \dd.}
\label{fig:erroranalysis}
\end{figure*}

\subsection{Micro Error Analysis: Per-Example Analysis}

Calibration plots provide a high-level summary of the quality of a KBC
system, from which developers can get an abstract sense of how to
improve it.  For developers to produce rules that can improve the
system, they need to actually look at data. We describe this
fine-grained error analysis, illustrated in
Figure~\ref{fig:erroranalysis}.

Figure~\ref{fig:erroranalysis}(a) illustrates the overall workflow of
error analysis. The process usually starts with the developer finding
a set of errors, which is a set of random variables having
predictions that are inconsistent with the training data. Then, for each random
variable, the developer looks at the corresponding tuples in the user
relation (recall that each tuple in the user relation corresponds to a
random variable) and tries to identify the types of errors. We
describe two broad classes of errors that we think of as improving the
recall and the precision, which we describe below.

\paragraph*{Recall Error Analysis.} 
A recall error, i.e., a false negative error, occurs when \dd is
expected to extract a fact but fails to do so.  In \dd's data model,
this usually means that a random variable has a low probability or
that the corresponding fact is not extracted as a candidate.
Figure~\ref{fig:erroranalysis}(b) illustrates the process of debugging
a recall error.

First, it is possible that the corresponding fact that is expected to
be extracted does not appear as a candidate. In this case, there is no
random variable associated with this fact, so it is impossible for \dd
to extract it. For example, this case might happen when the
corresponding fact appears in tables, but we only have a text-based
extractor. In this case, the developer needs to write extractors to
extract these facts as candidates.

Another common scenario is when the corresponding fact appears as a
candidate but is not assigned a high enough probability by \dd.  In
this case, the developer needs to take a series of steps, as shown in
Figure~\ref{fig:erroranalysis}(b), to correct the error.  First, it is
possible that this candidate is not associated with any features or
that the corresponding features have an obvious bug. In this case,
one needs to fix the extractors of the features. Second, it is
possible that the corresponding features have a relatively low weight
learned by \dd. In that case, one needs to look at the training data
to understand the reason. For example, one might want to add more
distant supervision rules to increase the number of positive
examples that share the same feature or even manually label more
positive examples. One might also want to change the feature
extractors to collapse the current features with other features to
reduce the sparsity.

\paragraph*{Precision Error Analysis.} 
A precision error, i.e., a false positive error, occurs when \dd outputs
a high probability for a fact, but it is an incorrect one or not
supported by the corresponding
document. Figure~\ref{fig:erroranalysis}(c) illustrates the process of
debugging a precision error. Similar to debugging recall errors, for
precision errors, the developer needs to look at the features
associated with the random variable. The central question to
investigate is why some of these features happen to have high
weights. In our experience, this usually means that we do not have
enough negative examples for training. If this is the case, the
developer adds more distant supervision rules or (less commonly)
manually labels more negative examples.

\paragraph*{Avoiding Overfitting.} 
One possible pitfall for the error analysis at the per-example level
is overfitting, in which the rules written by the developer overfit to
the corpus or the errors on which the error analysis is conducted. To
avoid this problem, we have insisted in all our applications on
conducting careful held-out sets to produce at least (1) a training set;
(2) a testing set; and (3) an error analysis set. In the error
analysis process, the developer is only allowed to look at examples in
the error analysis set, and validate the score on the testing set.

\section{Related Work} \label{sec:related_work}

{\em Knowledge Base Construction} (KBC) has been an area of intense
study over the last decade~\cite{Krishnamurthy:2009:SIGMODRec,
  Shen:2007:VLDB,Suchanek:2009:WWW,Nakashole:2011:WSDM,Etzioni:2004:WWW,
  Yates:2007:NAACL,Banko:2007:IJCAI,Kasneci:2009:SIGMODR,Betteridge:2009:AAAI,
  Carlson:2010:AAAI,Zhu:2009:WWW,Poon:2007:AAAI}. Within this space,
there are a number of approaches.

\paragraph*{Rule-Based Systems.} The earliest KBC systems used 
pattern matching to extract relationships from text. The most
well-known example is the ``Hearst Pattern'' proposed by
Hearst~\cite{Hearst:1992:COLING} in 1992. In her seminal work, Hearst
observed that a large number of hyponyms can be discovered by simple
patterns, e.g., ``X such as Y.'' Hearst's technique has formed
the basis of many further techniques that attempt to extract
high-quality patterns from text.  Rule-based (pattern matching-based)
KBC systems, such as IBM's
SystemT~\cite{Krishnamurthy:2009:SIGMODRec,Li:2011:ACL}, have been
built to aid developers in constructing high-quality patterns.  These
systems provide the user with a (declarative) interface to
specify a set of rules and patterns to derive relationships.  These
systems have achieved state-of-the-art quality on tasks such as
parsing~\cite{Li:2011:ACL}.

\paragraph*{Statistical Approaches.}
One limitation of rule-based systems is that the developer needs to
ensure that all rules provided to the system are high-precision
rules. For the last decade, probabilistic (or machine learning)
approaches have been proposed to allow the system to select from a
range of a priori features automatically. In these approaches, the
extracted tuple is associated with a marginal probability that it is
true. \dd, Google's knowledge graph, and IBM's Watson are built on
this approach. Within this space, there are three styles of
systems:

\begin{itemize}
\compactify 
\item \textbf{Classification-Based Frameworks.} Here, traditional classifiers assign each tuple a probability score,
  e.g., a na\"ive Bayes classifier or a logistic regression
  classifier. For example, KnowItAll~\cite{Etzioni:2004:WWW} and
  TextRunner~\cite{Yates:2007:NAACL,Banko:2007:IJCAI} use a na\"ive
  Bayes classifier, and CMU’s
  NELL~\cite{Betteridge:2009:AAAI,Carlson:2010:AAAI} uses logistic
  regression. Large-scale systems typically use these types of
  approaches in sophisticated combinations, e.g., NELL or Watson.

\item \textbf{Maximum a Posteriori (MAP).} Here, a
  probabilistic approach is used, but the MAP or most likely world
  (which do differ slightly) is selected. Notable examples include the YAGO
  system~\cite{Kasneci:2009:SIGMODR}, which uses a PageRank-based
  approach to assign a confidence score. Other examples include
  SOFIE~\cite{Suchanek:2009:WWW} and
  Prospera~\cite{Nakashole:2011:WSDM}, which 
  use an approach based on constraint satisfaction.

\item \textbf{Graphical Model Approaches.} The classification-based
  methods ignore the interaction among predictions, and there is a
  hypothesis that modeling these correlations yields higher quality
  systems more quickly. A
  generic graphical model has been used to model the probabilistic distribution
  among all possible extractions. For example, Poon et
  al.~\cite{Poon:2007:AAAI} used Markov logic networks
  (MLN)~\cite{Domingos:2009:Book} for information
  extraction. Microsoft's
  StatisticalSnowBall/EntityCube~\cite{Zhu:2009:WWW} also uses an
  MLN-based approach. A key challenge in these systems is
  scalability. For example, Poon et al. was limited to 1.5K
  citations. Our relational database-driven algorithms for
  MLN-based systems are dramatically more scalable~\cite{Niu:2011:VLDB}.
\end{itemize}

\section{Future Work}

\dd is our first step toward facilitating the building of knowledge
base construction systems of sufficient quality to support
applications, including rigorous scientific discoveries. Given our
experience in interacting with scientists and understanding their
needs, we found the following directions interesting candidates for
future exploration, most of which seem to be challenging open research
problems.

\begin{itemize}
\compactify

\item {\bf Dealing with Time and Temporal Information.} The KBC model
  we describe in this work treats time as a distinguished predicate
  such that each fact is extracted as long as there exists a time
  point such that it is true. However, prior work has shown that each
  fact could be associated with a more expressive temporal
  tag~\cite{Ling:2010:AAAI,Wang:2010:EDBT,Wang:2011:WWW}, and our
  previous participation in the DARPA machine reading challenge also
  attempted to produce a KBC system with such tags.  From our
  conversations with scientists, we have found that these temporal
  tags are sometimes important. For example, in paleontology, where
  the corpus spans more than 400 years, the ability to find the most
  recent version of the fact is important to produce the up-to-date
  view of the tree of life. It is interesting to study how to
  integrate temporal information into \dd's KBC model and conduct
  reasoning over it.

\item {\bf Visual Information Extraction.} As we mentioned in
  Section~\ref{sec:kbc}, we observed that there is a vast amount of
  information buried in data sources such as images and
  diagrams. Although the current \dd system can perform some simple
  image-based extraction, e.g., to extract body size from paleontology
  publications~\cite{Peters:2014:ArXiv}, it is an interesting
  direction to study how to extract information from tables, charts,
  and diagrams.

\item {\bf Incremental Processing.} One observation we have about
  debugging a KBC system is that, to conduct error-analysis
  efficiently, one usually needs to rerun the system with slightly
  different \dd programs (e.g., more rules) and data (e.g., more
  structural resources). In our application, it is not uncommon for
  the factor graph to contain billions of random variables. One
  obvious solution to this is to study how to incrementally conduct
  inference and learning to support more efficient error
  analysis. This problem is distinct from incremental or online
  learning, as the focus is on maintaining the downstream data
  products. We have done some work in this direction in a simplified
  classifier-based setting~\cite{DBLP:journals/pvldb/KocR11}.

\item {\bf ``Active Debugging'' and Rule Learning.} The current
  debugging infrastructure in \dd is ``passive'' in that \dd waits for
  instructions from the user for error analysis or adding more
  features. One future direction is to study whether \dd can play a more
  ``active'' role.  One example is whether \dd could automatically
  suggest errors for the users to investigate or report relations
  where more training data are required. It is also interesting to
  study whether it is possible for \dd to automatically learn
  inference rules or feature extractors and recommend them to
  the user. We envisioned a similar system in our previous
  work~\cite{Anderson:2013:CIDR} and have worked on a domain-specific
  language for a related problem~\cite{Zhang:2014:SIGMOD}.

\item {\bf Even-More-joint Information Extraction.} As we show in this
  work, one of \dd's advantage is the ability to jointly take
  advantage of different sources of signals. One interesting question
  is whether we can extend the current KBC model to be even more joint. First, it is interesting to study how to use \dd for
  low-level NLP tasks, e.g., linguistic parsing and optical character
  recognition, and how these low-level tasks interact with high-level
  tasks, e.g., relation extraction. Second, it is interesting to study
  how application-level knowledge, e.g., the R-script used by
  paleontologists over \dd's knowledge base, can be used to constrain
  and improve the extraction task.

%% \item {\bf Integrating Human Feedback in Reliable ways.} As we argue
%%   in this paper, developing a high-quality system is often an
%%   interactive human-in-the-loop effort. That is, \dd needs to take
%%   feedback, in different levels, from human, with different
%%   professions and training. Thus, one interesting question is how to
%%   make this feedback as reliable as possible~\cite{Zhang:2012:ACL}.

\item {\bf Visualization of Uncertainty.} The current \dd system
  expresses the uncertainty of inference results to the user using
  marginal probabilities. Even simple visualizations have proved to be
  far more effective at conveying debugging information about
  statistical or aggregate properties.

\end{itemize}

\section{Conclusion}

We have described what we believe is one of the key features of \dd:
the ability to rapidly debug and engineer a better system. We have
argued that probabilistic and machine learning techniques are critical,
but only in that they enable developers to think in terms of
features---not algorithms.

% \section{Discussion and Future Work} \label{sec:future}
% \paragraph*{Deeper NLP Understanding}
% Feiran (Parser)
% Zifei (OCR)
% \paragraph*{Beyond Text Processing}
% Amir (Vision)
% \paragraph*{Evolving Corpus and Program}
% \paragraph*{Feature Recommendation and Rule Learning}

\section{Acknowledgments}

We would like to thank the users of DeepDive, especially Shanan Peters
and Emily Doughty, who have given a great deal of helpful (and
patient) feedback. We would also like to thank Michael J. Cafarella
and Dan Suicu who gave comments on an earlier draft of this document.
We gratefully acknowledge the support of the Defense Advanced Research
Projects Agency (DARPA) XDATA Program under No. FA8750-12-2-0335 and
DEFT Program under No. FA8750-13-2-0039, DARPA's MEMEX program, the
National Science Foundation (NSF) CAREER Award under No. IIS-1353606
and EarthCube Award under No. ACI-1343760, the Office of Naval
Research (ONR) under awards No. N000141210041 and No. N000141310129,
the Sloan Research Fellowship, American Family Insurance, Google, and
Toshiba.  Any opinions, findings, and conclusions or recommendations
expressed in this material are those of the authors and do not
necessarily reflect the views of DARPA, AFRL, NSF, ONR, or the
U.S. government.

%\newpage
\small
{\bibliographystyle{abbrv} \bibliography{ieeedeb}}

\end{document}